# Predictive Dosimetry in PSMA-Targeted Radiopharmaceutical Therapies: A PBPK Modeling and Machine Learning Study

Hamid Abdollahi, James Fowler, Carlos Uribe, Arman Rahmim

***Abstract*** **Predictive dosimetry is central to enabling personalized radiopharmaceutical therapy (RPT), particularly in prostate-specific membrane antigen (PSMA)–targeted theranostics. In this work, we develop a three-layer computational framework that integrates physiologically based pharmacokinetic (PBPK) modeling with machine learning (ML) to predict both physical (AUC, absorbed dose) and biological (BED, EQD2) dosimetric endpoints in tumors and major organs. In the first layer, we generated 640 virtual patients using PBPK simulations of $^{18}$F-, $^{68}$Ga-, and $^{64}$Cu-labeled PSMA PET tracers paired with $^{177}$Lu-PSMA therapy, producing 15,360 tumor and organ time–activity curves (TACs) under realistic biological variability and PET-like noise. In the second layer, TACs were transformed into quantitative kinetic features and mapped to physical and biological dose metrics. In the third layer, ML models (Random Forest, Extra Trees, Ridge, Gradient Boosting, and XGBoost) were trained to predict RPT doses from PET-derived features, with performance evaluated using mean absolute percentage error (MAPE) and $R^2$. $^{64}$Cu-PSMA-617–based PET yielded the most robust predictions, achieving tumor dose MAPE as low as 8% and 10–20% for normal organs, while $^{18}$F-DCFPyL showed volume-dependent performance and $^{68}$Ga-PSMA-11 exhibited higher variability. SHAP analysis revealed that peak uptake (AMax), clearance, and early kinetic features dominated predictive performance across organs and endpoints. This PBPK-ML framework enables scalable, physiology-informed predictive dosimetry and provides a powerful foundation for trial design and patient-specific treatment planning in PSMA-targeted RPT. These results demonstrate that pre-therapy PET can serve as a reliable surrogate for post-therapy dosimetry, enabling scalable personalization of PSMA-targeted RPT.**

## I. INTRODUCTION

Nuclear medicine is undergoing a major renaissance driven by the rapid emergence of radiopharmaceutical therapies (RPTs), which combine molecular targeting with precision radiation delivery to enable biologically guided cancer treatment [1]. Modern RPTs employ radionuclides such as $^{177}$Lu and $^{225}$Ac to deliver cytotoxic radiation doses directly to tumor cells while minimizing exposure to surrounding healthy tissues, thereby supporting more personalized and effective therapy [2]. This paradigm is further strengthened by theranostic strategies that integrate diagnostic imaging and therapy using the same molecular targeting vectors, allowing patient selection, treatment planning, and response assessment within a unified framework [3].

Among the most impactful theranostic applications are prostate-specific membrane antigen (PSMA)–targeted radiopharmaceuticals for the management of metastatic castration-resistant prostate cancer (mCRPC) [4]. In this setting, PET imaging agents labeled with radionuclides such as $^{18}$F, $^{68}$Ga, and $^{64}$Cu enable sensitive detection and quantification of PSMA-expressing tumors [5]. Once sufficient PSMA expression is confirmed, therapeutic counterparts—typically labeled with $^{177}$Lu or $^{225}$Ac—are administered to deliver targeted radiation to malignant tissue. A wide range of synthetic PSMA ligands allows flexible pairing between diagnostic and therapeutic radionuclides, supporting theranostic optimization according to clinical need [6]. In this framework, tumor targeting is governed by the biological interaction between the vector and PSMA, while the radionuclide functions solely as an imaging or therapeutic payload.

Despite the rapid clinical adoption of PSMA-targeted RPTs, treatment is still largely administered using fixed or weight-based activities rather than patient-specific predictive dosimetry. In mCRPC, this practice fails to account for large inter-patient variability in tumor burden, PSMA expression, organ uptake, and clearance kinetics, leading to potential under-treatment in some patients and excessive normal-tissue irradiation in others [7,8]. Although landmark trials such as VISION demonstrated improved overall survival and progression-free survival using $^{177}$Lu-PSMA-617, treatment was delivered using standardized activity prescriptions rather than individualized dose planning [9]. Consequently, personalized dosimetry—where administered activity is tailored to predicted absorbed doses in tumors and organs at risk—remains a critical unmet need in PSMA-targeted therapy [10]. Emerging trial designs increasingly aim to integrate imaging biomarkers, radiobiological modeling, and artificial intelligence (AI)–based prediction to optimize therapeutic efficacy and safety on an individual basis [11].

In parallel, precision oncology has seen growing interest in computational modeling frameworks that integrate biological, physiological, and imaging data to predict therapeutic outcomes and guide treatment optimization [12–14]. In radiopharmaceutical therapy, theranostic digital twins (TDTs) have been proposed as computational surrogates that represent patient-specific physiology and tumor biology, allowing *in silico* simulation of radiopharmaceutical distribution, uptake, and clearance before treatment delivery [5,16]. Building on this concept, virtual clinical trials (VCTs) provide a powerful means to explore therapeutic response, toxicity, and protocol design across large populations of simulated patients, thereby enabling hypothesis testing and trial optimization prior to clinical implementation [17–19].

Physiologically based pharmacokinetic (PBPK) modeling has emerged as a cornerstone of such in silico frameworks for RPTs [20–23]. PBPK models describe tracer transport, receptor binding, internalization, and clearance using mechanistic representations of physiology, enabling simulation of time–activity curves (TACs) in tumors and organs under varying biological conditions [24]. These models support evaluation of dose delivery, injection schedules, and tracer selection, and can be used to prioritize radiopharmaceutical candidates for clinical translation before committing resources to clinical trials. Such efforts are part of a broader shift toward in silico clinical trials [25], which is increasingly recognized by regulatory agencies. Notably, the FDA and the European Union have both endorsed computational modeling in drug development, and recent U.S. legislation explicitly permits investigational drugs to advance to human trials based on non-clinical computer-based evidence [26,27].

PBPK models also play a critical role in explaining why patients receiving the same injected activity may experience very different absorbed doses and clinical outcomes [28,29]. By incorporating patient-specific parameters such as tumor PSMA expression, blood flow, renal function, and receptor density, PBPK-based simulations enable individualized prediction of biodistribution and dose delivery [20,30,31]. This mechanistic foundation is essential for moving beyond empirical activity prescription toward truly personalized radiopharmaceutical therapy.

While pre-therapy PET imaging has been used to estimate therapeutic dose using metrics such as standardized uptake value (SUV), volumetric indices, and radiomic features [32,33], most existing approaches rely on static or semi-quantitative measures that fail to capture the dynamic pharmacokinetics governing radiopharmaceutical transport and retention. To date, no study has fully combined PBPK-guided dynamic PET simulation with machine-learning–based modeling to predict patient-specific absorbed dose in large virtual populations.

In this study, we therefore develop a PBPK-ML–based predictive dosimetry framework that links PET-derived TAC features to physical and biological dose endpoints across thousands of virtual PSMA-targeted therapy scenarios. By integrating mechanistic pharmacokinetic modeling with data-driven learning, our approach provides a scalable platform for pre-therapy dose prediction, protocol optimization, and in silico trial design for PSMA-targeted RPT in mCRPC and potentially other cancers.

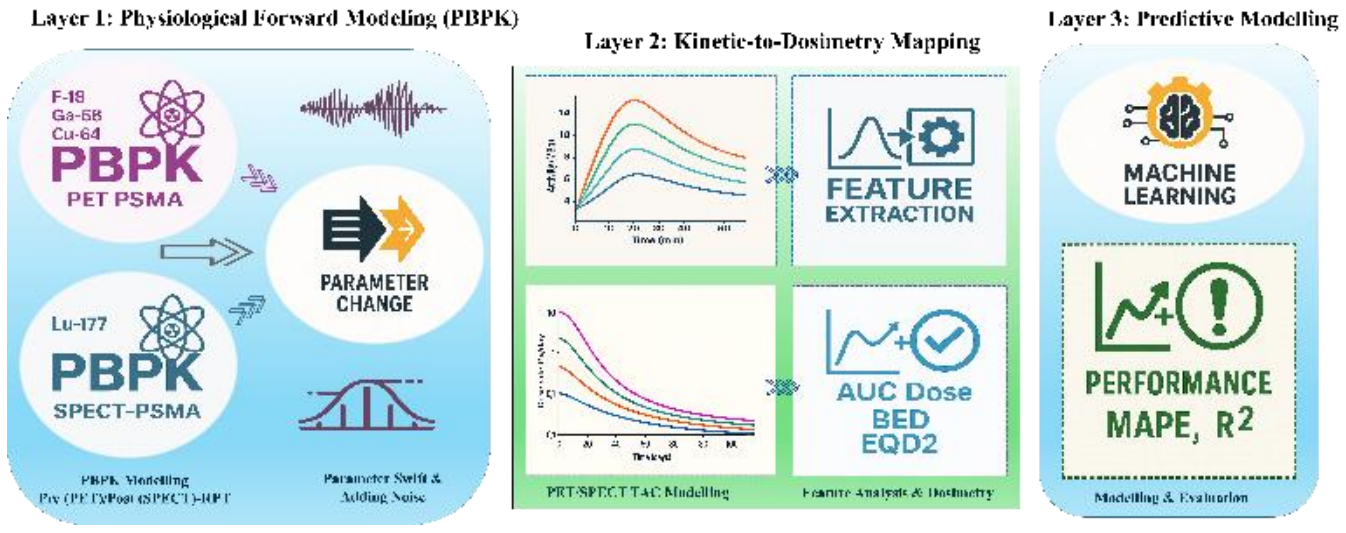

Fig. 1. Three-layer PBPK–AI framework for predictive dosimetry. PBPK simulation generates PET and RPT TACs, kinetic features are mapped to AUC, dose, BED, and EQD2, and machine learning predicts therapy dosimetry from pre-therapy PET with performance evaluated by MAPE and $R^2$.

## II. METHODS

The overall workflow is summarized in Fig. 1. The framework is organized into three coupled modeling layers: (i) PBPK-based physiological simulation, (ii) kinetic-to-dosimetry mapping, and (iii) machine-learning–based predictive modeling and evaluation. Layer 1 generates physiologically realistic time–activity curves (TACs) for PET and RPT radiopharmaceuticals. Layer 2 converts these TACs into kinetic descriptors and physical and biological dose quantities. Layer 3 learns the mapping from PET-derived kinetics to post-therapy dosimetry.

### *A. Layer 1: PBPK-Based Physiological Simulation*

### *A.1 PBPK Modeling Framework and Validation*

PBPK modeling was used as the mechanistic backbone of the framework to simulate PSMA-targeted radiopharmaceutical transport, binding, internalization, and clearance at the organ and lesion level. The PBPK model represents PSMA-mediated vector biology, rather than the radionuclide itself, and supports multiple imaging and therapy tracers by coupling the same biological model to radionuclide-labeled PSMA ligands, including $^{18}$F-DCFPyL, $^{68}$Ga-PSMA-11, $^{64}$Cu-PSMA-617, and $^{177}$Lu-PSMA-I&T [16, 31].

The model extends our previously validated PSMA-PBPK framework [20] and represents each organ and tumor using four physiologically distinct sub-compartments: vascular, interstitial, receptor-bound, and intracellular. Ligand transport is governed by blood flow, capillary permeability, and receptor-mediated binding and internalization. Binding is described by association and dissociation rate constants $k_{\mathrm{on}}$and $k_{\mathrm{off}}$, while intracellular trafficking is governed by the internalization rate $\lambda_{\mathrm{int}}$and release rate $\lambda_{\mathrm{release}}$. These processes are implemented through a reaction-network formalism in MATLAB SimBiology, allowing simultaneous simulation of free ligand, receptor-bound complex, and albumin-bound fractions without parallel kinetic tracks.

Tumors were modeled as physiologically distinct compartments with explicit volume dependence and heterogeneous receptor expression. To represent intra-tumoral heterogeneity, each tumor was divided into a high-PSMA sub-compartment and a remainder compartment using a tumor flow fraction $f_{\mathrm{TU},1}$(Table I). Normal organs including kidneys, salivary glands, liver, spleen, and bone marrow were modeled using the same compartmental structure with organ-specific blood flow, receptor density, and clearance parameters.

Physiological scaling was applied on a per-patient basis using body height (BH) and body surface area (BSA) to determine organ volumes and blood flows via standard allometric relationships [34]. Renal handling was governed by the tubular extraction ratio (TER) and urinary excretion fraction $f_{\mathrm{ex}}$, with TER varied across the clinically validated range 0.2–0.4 $\mathrm{L \cdot min^{-1}}$ based on MAG3 renal clearance data [34], [40]. All kinetic, anatomical, and clearance parameters were sampled within physiologically explicit bounds defined in Table I using experimental and clinical literature [34, 40].

The PBPK model was validated against the experimental PSMA PET and therapy data of Kletting *et al.* [34], ensuring physiologically realistic uptake, clearance, and tumor-to-organ ratios. The full model is provided in SBML format to allow reproducibility and independent verification.

TABLE I. PBPK MODEL PARAMETERS, RANGES, AND LITERATURE SOURCES USED TO SIMULATE PSMA TRACER KINETICS ACROSS TUMOR AND NORMAL ORGANS.

| Parameter | Symbol | Value / Range | Unit | Definition / Role in Model | Source |
|---|---|---|---|---|---|
| Tumor flow fraction | $f_{\mathrm{TU},1}$ | 0.5 | – | Fraction of total tumor volume assigned to Tumor1; models intra-tumoral heterogeneity | [34] |
| Association rate | $k_{\mathrm{on}}$ | 0.046 | L·nmol⁻¹·min⁻¹ | Ligand–receptor association rate | [34,35] |
| Dissociation constant | $K_D$ | 1–8 | nmol·L⁻¹ | Ligand–receptor binding affinity | [34–36] |
| Dissociation rate | $k_{\mathrm{off}} = K_D \cdot k_{\mathrm{on}}$ | 0.046–0.368 | min⁻¹ | Rate of ligand detachment from receptor | [34] |
| Internalization rate | $\lambda_{\mathrm{int}}$ | 0.001 | min⁻¹ | Internalization of ligand–receptor complex | [34,37] |
| Release rate | $\lambda_{\mathrm{release}}$ | 0.0003–0.003 | min⁻¹ | Release of internalized tracer to extracellular space | [34] |
| Tumor receptor density | $[R_{\mathrm{TU},0}]$ | 50–3500 | nmol·L⁻¹ | PSMA expression level in tumor | [34] |
| Tumor remainder receptor density | $[R_{\mathrm{TU,rest},0}]$ | 266 | nmol·L⁻¹ | Baseline PSMA density in tumor remainder | [34] |
| Kidney receptor density | $[R_{\mathrm{K},0}]$ | 20–200 | nmol·L⁻¹ | PSMA expression in kidneys | [34] |
| Salivary gland density | $[R_{\mathrm{SAL},0}]$ | 30–300 | nmol·L⁻¹ | PSMA expression in salivary glands | [34,38] |
| Tumor perfusion density | $f_{\mathrm{TU}}$ | 0.5 | mL·min⁻¹·g⁻¹ | Tumor blood flow density | [34,39] |
| Tubular extraction ratio | TER | 0.2–0.4 | L·min⁻¹ | Active renal extraction ($^{99m}$Tc-MAG3 based) | ]34,40] |
| Excretion fraction | $f_{\mathrm{ex}}$ | 0.96 | – | Fraction of filtered tracer excreted | [34] |
| Body height | BH | 150–200 | cm | Used for physiological scaling | ]34] |
| Body surface area | BSA | 1.5–2.2 | m² | Calculated from BH and BW | ]34 |

TABLE II. KEY FEATURES EXTRACTED FROM TIME-ACTIVITY CURVES (TACS)

| Feature Name | Definition | Feature Set Category |
|---|---|---|
| Tmax | Time at which maximum activity occurs | Temporal Features |
| AMax | Maximum activity value in MBq | Amplitude & Intensity |
| Time To Peak | Time from start to Tmax | Temporal Features |
| Tmin | Time at which minimum activity occurs | Temporal Features |
| AMin | Minimum activity value | Amplitude & Intensity |
| Time From Peak To Min | Time difference from peak to minimum | Temporal Features |
| AMean | Mean activity across the TAC | Amplitude & Intensity |
| AStdev | Standard deviation of activity values | Amplitude & Intensity |
| AMedian | Median activity value | Amplitude & Intensity |
| Skewness | Measure of asymmetry in the TAC distribution | Distribution & Shape |
| Kurtosis | Sharpness or flatness of the TAC peak | Distribution & Shape |
| Entropy | Shannon entropy; quantifies signal complexity | Distribution & Shape |
| Energy | Total energy (sum of squared activities) | Distribution & Shape |
| AUC | Area under the TAC using Simpson's rule | Amplitude & Intensity |
| Half Life | Estimated time to reach half of AMax post-peak | Temporal Features |
| Clearance | Ratio of AMax to AUC; reflects washout behavior | Amplitude & Intensity |
| P1–P99 | Percentile values (1st to 99th) of TAC activity | Percentile-Based |
| T10–T100 | Time points reaching ≥ X% of AMax during the rising phase | Temporal Features |
| T-90 to T-10 | Time points post-peak where activity falls to ≤ X% of AMax | Temporal Features |
| Increasing Slope | Maximum positive slope of TAC segments | Slope & Kinetics |
| Decreasing Slope | Maximum negative slope of TAC segments | Slope & Kinetics |
| Max Slope / MinSlope | Overall max/min slope from the TAC | Slope & Kinetics |

*A.2 Generation of Virtual Patient Cohorts*

Virtual patient populations were generated by systematically varying PBPK parameters that influence PSMA tracer kinetics and radiation dose delivery. These included receptor densities, tumor volumes, binding kinetics, perfusion, renal clearance, and physiological scaling parameters. Tumor volumes of 0.001, 0.01, 0.1, and 1 L were simulated, and three biological scenarios were defined by increasing PSMA receptor densities in tumors and normal organs, representing low-, medium-, and high-expression phenotypes.

To account for radiopharmaceutical-specific physicochemical differences among PET and therapy agents, all relevant PBPK parameters were normalized by the relative molecular weights of $^{18}$F-DCFPyL, $^{68}$Ga-PSMA-11, and $^{64}$Cu-PSMA-617 with respect to $^{177}$Lu-PSMA-I&T. To represent intra- and inter-patient biological variabilities, each PBPK parameter was further perturbed using:

$$\text{Parameter}_{\text{noisy}} = \text{Parameter}_{\text{original}}(1 + 0.2\,N(0,1)) \quad (1)$$

where $N(0,1)$ is a standard normal random variable. Using these parameter sets, full TACs were generated for PET tracers and for $^{177}$Lu therapy over 30,000 min (≈500 h), capturing complete biological clearance and physical decay.

### *A.3 PET-Like Noise Modeling*

PET noise was applied directly to PET-side TACs, while RPT-side TACs were kept noise-free to preserve dosimetric ground truth. EM-based PET reconstruction induces approximately log-normal voxel statistics, while at the ROI-level statistics are approximately normal, and noise is proportional to square-root of mean uptake [41, 42] and as such was modeled as

$$\sigma = \alpha\sqrt{\mu}, \quad (2)$$

with $\alpha = 0.001$.

## *B. Layer 2: Kinetic-to-Dosimetry Mapping*

Layer 2 converts PBPK-generated TACs into PET-like kinetic descriptors and RPT-based dosimetry targets.

### *B.1 Feature Extraction from PET TACs*

PET TACs were converted to activity units and processed to extract kinetic descriptors summarizing uptake magnitude, timing, washout, shape, and slope behavior. The extracted features, defined in Table II, include $T_{\max}$, $A_{\max}$, AUC, clearance $A_{\max}/\text{AUC}$, entropy, skewness, kurtosis, percentile statistics, and slope-based measures. These features form the ML input space.

### *B.2 Dosimetry from RPT TACs*

For $^{177}$Lu TACs, cumulative exposure was computed as

$$\text{AUC} = \int_0^{t_n} N_i(t)\,dt, \quad (3)$$

with $t_n = 30{,}000$min. Dose rate and absorbed dose were computed as

$$\dot{D}_i(t) = N_i(t)\lambda S_{i\leftarrow i}, \quad (4)$$

$$D = \int_0^{t_n} \dot{D}_i(t)\,dt. \quad (5)$$

BED and EQD2 were computed using

$$G(T) = \frac{2}{D^2}\int_0^T \dot{D}(t)\int_0^t \dot{D}(w)e^{-\mu(t-w)}dw\,dt, \quad (6)$$

$$\text{BED} = D\left(1+\frac{GD}{\alpha/\beta}\right), \quad (7a)$$

$$\text{EQD2} = \frac{\text{BED}}{1+2/(\alpha/\beta)}. \quad (7b)$$

## *C. Layer 3: Machine-Learning–Based Predictive Modeling and Performance Evaluation*

### *C.1 Learning Task Definition*

Layer 3 establishes a supervised regression mapping from pre-therapy PET-side kinetics (Layer 2) to post-therapy dosimetry endpoints computed from $^{177}$Lu-PSMA TACs (Layer 2). For each virtual patient $i$, let $\mathrm{x}_i \in \mathbb{R}^p$denote the vector of TAC-derived kinetic features extracted from the PET-side TACs (Table II), and let $y_i$denote a target dosimetric endpoint computed from the therapy-side TACs. Separate regression models were trained for each endpoint $y \in \{\text{AUC}, D, \text{BED}, \text{EQD2}\}$and for each structure (tumor and normal organs). The learning objective is to estimate a function $f(\cdot$such that

$$\hat{y}_i = f(\mathrm{x}_i), \quad (9)$$

where $\hat{y}_i$is the predicted dosimetric quantity for patient $i$. This formulation explicitly reflects the predictive dosimetry paradigm of inferring post-therapy dose surrogates using pre-therapy tracer kinetics, while recognizing that in the present study TACs are PBPK-generated and noise-augmented rather than extracted from reconstructed images.

### *C.2 Cross-Validation Design and Data Partitioning*

Model performance was evaluated using patient-level five-fold cross-validation to ensure strict separation between training and testing data and to prevent information leakage. Specifically, all TACs and derived features belonging to a given virtual patient were assigned to the same fold, such that no patient contributed data to both training and test sets in any fold. For fold $k$, the regression model was trained on patients in the remaining four folds and tested on the held-out patients in fold $k$. Performance metrics were computed on the held-out fold and then averaged across folds to obtain robust estimates of generalization performance.

Because the data were generated under both No-Noise and Noise conditions (Layer 2), the complete cross-validation procedure was performed independently for each condition, enabling separation of intrinsic algorithmic performance (No-Noise) from robustness to PET-like TAC variability (Noise).

### *C.3 Feature Conditioning and Dimensionality Control*

All preprocessing steps were performed *within each training fold only* and then applied to the corresponding test fold using parameters derived from training data to avoid leakage. Features with more than 20% missing values were excluded prior to model fitting. For features with remaining missing values below this threshold, imputation was performed using the median value computed from the training fold for each feature, and the same imputation values were applied to the test fold.

To minimize redundancy and instability due to collinearity, highly correlated features were removed using Pearson's correlation computed within the training fold. If two features had $|r| > 0.90$, one was removed based on lower univariate association with the target within the training fold. After filtering, all retained features were standardized using z-score normalization based on the training-fold mean and standard deviation:

$$x'_{ij} = \frac{x_{ij} - \mu_j}{\sigma_j}, \quad (10)$$

where $\mu_j$and $\sigma_j$are the mean and standard deviation of feature $j$computed from the training fold.

### C.4 Feature Selection and Stability Analysis

To improve interpretability and reduce overfitting risk in high-dimensional kinetic feature spaces, feature selection was performed independently within each training fold using three complementary approaches: recursive feature elimination (RFE), LASSO regression with cross-validated $L_1$regularization, and Random-Forest–based feature-importance filtering. For each approach, selection was performed using only the training-fold data, and the selected feature subset was then used to train the downstream regression model, which was subsequently evaluated on the held-out test fold.

A fixed number of $K = 5$features was retained per fold. This design choice was adopted to balance model stability,

interpretability, and the risk of overfitting given the effective sample size per fold and the strong inter-feature correlations typical of TAC-derived descriptors. The selected features and their frequencies across folds were tracked to characterize selection consistency.

To quantify feature-selection robustness beyond fold-to-fold variability, stability selection was performed for the No-Noise dataset across all four targets. In each of 50 iterations, a random 50% subset of virtual patients was sampled (without replacement), feature selection was rerun using the same pipeline, and the selected feature identities were recorded. For feature $j$, the stability score was defined as

$$\text{Stability}_j = 100 \times \frac{s_j}{50}, \qquad (11)$$

where $s_j$ is the number of iterations in which feature $j$was selected. Features exceeding 40% stability were considered robust and were reported as reproducible kinetic signatures for the corresponding endpoint and organ.

*C.5 Regression Models and Training Strategy*

Five classical regression models were evaluated for prediction of $\{\text{AUC}, D, \text{BED}, \text{EQD2}\}$from the selected TAC features: Random Forest (200 trees), Extra Trees (400 trees), Ridge regression ($\alpha = 1.0$), Gradient Boosting, and Support Vector Regression (SVR; $C = 1.0$, $\epsilon = 0.2$). For each cross-validation fold and each endpoint, a fresh model instance was trained to ensure independence between folds.

Because high-dose (and high-BED/EQD2) samples are often underrepresented but clinically influential, two complementary strategies were used to emphasize these cases during training. First, targeted oversampling was performed within each training fold by duplicating samples whose log-transformed targets exceeded the 75th percentile of the training distribution. Second, a nonlinear sample-weighting function was applied during training:

$$w(y) = 1 + \frac{\log(y) - \mu}{\sigma}, \qquad (12)$$

where $y$is the target value (Dose, BED, or EQD2), and $\mu$and $\sigma$are the mean and standard deviation of $\log(y)$computed from the training fold. These weights were used directly in model fitting for algorithms supporting sample weights (e.g., tree-based ensemble methods and boosting), while for models without native weighting support the oversampling strategy provided an approximate emphasis on high-target regimes. Model performance obtained using nonlinear weighting was compared against unweighted training and linearly weighted alternatives to quantify the impact of emphasizing high-dose outcomes.

*C.6 Performance Metrics and Primary Endpoint*

Predictive performance was quantified using multiple regression metrics computed on held-out folds, including mean absolute error (MAE), mean squared error (MSE), root mean squared error (RMSE), median absolute error (MedAE), mean absolute percentage error (MAPE), and the coefficient of determination ($R^2$). For target $y$and predictions $\hat{y}$, MAPE was computed as

$$\text{MAPE} = \frac{1}{n}\sum_{i=1}^{n}\left|\frac{y_i - \hat{y}_i}{y_i}\right| \times 100\%, \qquad (13)$$

and was pre-specified as the primary endpoint because it is scale-independent and facilitates direct comparison across organs and tumor volumes. Metrics were averaged across the five cross-validation folds to summarize overall performance per model, endpoint, tracer, and noise condition.

To characterize bias and error structure, predicted–observed scatter plots, residual distributions, and kernel density estimates were generated for the top-performing models, enabling evaluation of systematic under- or over-estimation across the clinically relevant target range.

*C.7 Model Interpretability Using SHAP*

To improve transparency and clinical interpretability of the learned kinetic-to-dose mapping, SHAP (SHapley Additive exPlanations) analysis was applied to tree-based regressors trained on standardized features. For a given model output $\hat{y}$, SHAP decomposes predictions into additive contributions

$$\hat{y}_i = \phi_0 + \sum_{j=1}^{K}\phi_{ij}, \qquad (14)$$

where $\phi_0$is the expected prediction over the training data and $\phi_{ij}$denotes the contribution of feature $j$for sample $i$. SHAP TreeExplainer was used for Random Forest, Extra Trees, and Gradient Boosting models, with check_additivity = Falseto accommodate small numerical deviations in ensemble approximations. Beeswarm plots were generated to visualize global feature importance and to identify whether higher or lower kinetic feature values drove predictions toward higher absorbed dose, BED, or EQD2, thereby linking interpretable TAC signatures to dosimetry outcomes.

## III. RESULTS

*A. Generated Virtual Cases and PBPK-Based Dosimetry*

Using the PBPK framework described in Section II, a total of 640 virtual patients were generated by sampling physiologically realistic parameter variations. This resulted in 15,360 time–activity curves (TACs) spanning tumors and five organs (kidneys, salivary glands, liver, spleen, and bone marrow) under both noise-free and noise-augmented PET conditions, as well as corresponding RPT TACs. Specifically, the dataset comprised 11,520 PET-derived TACs $[640 \times 6 \text{ regions} \times 3 \text{ PET conditions}]$ and 3,840 RPT TACs $[640 \times 6 \text{ regions}]$.

Dosimetric outcomes derived from the RPT TACs are summarized in Table III for tumor volumes ranging from 0.001 L to 1 L. As tumor volume increased, tumor AUC, Dose, BED, and EQD2 rose substantially, while corresponding organ values remained relatively stable, resulting in a pronounced decline in organ-to-tumor ratios (OTRs). Tumor AUC increased from $0.03 \pm 0.03 \times 10^6$nmol at 0.001 L to $21.6 \pm 14.0 \times 10^6$nmol at 1 L. In contrast, the kidneys and liver exhibited the highest organ AUCs at small tumor volumes, reaching $0.96 \pm 0.86 \times 10^6$nmol and $0.91 \pm 0.55 \times 10^6$nmol, respectively.

Consequently, the kidney OTR declined sharply from 27.7 at 0.001 L to 0.029 at 1 L, consistent with a tumor sink effect and improved tumor selectivity at larger volumes. Tumor absorbed dose peaked at $48.3 \pm 38.1$ Gy for a tumor volume of 0.01 L and decreased to $31.6 \pm 20.4$ Gy at 1 L, whereas organ doses remained low across all volumes. The kidney received the highest organ dose ($\sim 4.67 \pm 4.12$ Gy), while spleen and bone marrow doses remained below 0.1 Gy. Similar trends were observed for BED and EQD2, with tumor BED ranging from $58.4 \pm 57.0$ Gy to $36.5 \pm 26.7$ Gy and tumor EQD2 decreasing from $40.6 \pm 38.8$ Gy to $24.1 \pm 17.6$ Gy as tumor volume increased. These patterns demonstrate physiologically consistent dosimetric behavior across tumor sizes and organs, validating the PBPK model as a suitable foundation for predictive dosimetry.

*B. Feature Correlation and Selection*

The relationships between pre-therapy PET-derived TAC features and dosimetric endpoints are illustrated in Fig. 2, which presents a comprehensive Pearson correlation heatmap. Strong correlations were observed among the dosimetric endpoints themselves ($r > 0.98$), reflecting their shared physical and radiobiological origins. Among all features, peak uptake (AMax) showed the strongest associations with outcomes ($r = 0.996$ with AUC and $r > 0.96$ with Dose, BED, and EQD2), identifying it as the dominant predictor across endpoints.

Additional activity-based features (AMean, AMedian, AStdev) and kinetic slope descriptors (e.g., MaxSlope) also exhibited strong correlations, emphasizing the importance of uptake intensity and temporal dynamics. Temporal percentiles (T10–T100) and activity percentiles (P10–P95) formed highly correlated clusters ($r > 0.9$), indicating substantial redundancy. In contrast, distributional features such as Skewness, Kurtosis, and Entropy showed weaker or negative correlations, suggesting that they encode complementary information related to TAC morphology rather than absolute uptake.

Feature selection was performed independently for each organ and endpoint after removal of highly correlated features ($|r| > 0.90$). Stability analysis across folds revealed consistent retention of a small subset of predictors, particularly for tumor and kidney models. The most frequently selected features across all conditions are summarized in Fig. 3, where AMax consistently ranked as the most influential feature. Clearance, Tmax, and P25 also appeared prominently, indicating the combined importance of peak uptake, washout behavior, and early kinetics. While AUC prediction additionally relied on distribution-based descriptors, Dose and BED emphasized early uptake features, and EQD2 showed increased sensitivity to both intensity- and shape-based metrics.

*C. Prediction Performance*

Prediction performance was evaluated using MAE, MSE, RMSE, MedAE, MAPE, and $R^2$, with MAPE used as the primary comparative metric. Tumor-level predictive performance across endpoints, radiopharmaceuticals, tumor volumes, and noise conditions is shown in Fig. 4. Overall, accuracy improved with increasing tumor volume, as reflected by narrower error distributions and lower median MAPE, while the introduction of noise consistently degraded performance, particularly at small volumes (0.001–0.01 L).

For AUC prediction, $^{64}$Cu-PSMA-617 demonstrated the most accurate and robust performance, with MAPE decreasing from approximately 5% at 0.001 L to about 2% at 1 L under noise, closely matching no-noise behavior. $^{18}$F-DCFPyL exhibited greater noise sensitivity at small volumes (MAPE > 6% at 0.001 L) but converged toward similar accuracy at larger volumes. In contrast, $^{68}$Ga-PSMA-11 showed broader and more overlapping error distributions, with relatively flat MAPE profiles across volumes, indicating reduced discriminative power under noisy conditions.

Dose prediction revealed stronger noise sensitivity, particularly for $^{18}$F-DCFPyL, where MAPE exceeded 20% at 0.001 L under noise and declined to approximately 12% at 1 L. $^{68}$Ga-PSMA-11 exhibited wide variability at intermediate volumes, whereas $^{64}$Cu-PSMA-617 maintained tighter distributions across all volumes, with MAPE decreasing from roughly 12% to 8%. BED and EQD2 followed similar trends, with $^{64}$Cu-PSMA-617 consistently yielding the lowest errors and strongest noise robustness, while $^{68}$Ga-PSMA-11 showed persistent instability, particularly for biologically weighted endpoints.

Predicted-versus-true dose scatter plots under noisy conditions are shown in Fig. 5. At small tumor volumes, noise led to increased scatter and systematic underestimation, especially for $^{18}$F-DCFPyL and $^{68}$Ga-PSMA-11, with absolute errors exceeding 100–150 Gy in some cases. In contrast, $^{64}$Cu-PSMA-617 maintained tighter clustering around the identity line, particularly for tumor volumes ≥ 0.1 L, where predictions became clinically reliable across all tracers.

Model-dependent performance trends for dose prediction are summarized in Fig. 6, which compares MAPE distributions across five regression models. Regularized and boosted models (Ridge and XGBoost) consistently achieved lower median errors and reduced variability, particularly for $^{64}$Cu-PSMA-617, while Random Forest and Extra Trees showed broader distributions under noise. Across all tracers, predictive accuracy improved with increasing tumor volume, underscoring the mitigating effect of uptake statistics on noise sensitivity.

Normal-organ prediction performance is summarized in Table III. Although noise increased errors overall, $^{64}$Cu-PSMA-617 consistently achieved the lowest MAPE across organs and endpoints, with particularly notable improvements for salivary gland Dose and red marrow EQD2. In contrast, $^{68}$Ga-PSMA-11 exhibited high baseline errors and substantial noise-induced variability, especially in low-uptake organs such as spleen and bone marrow.

*D. Model Interpretability*

To interpret the learned kinetic–dosimetric relationships, SHAP analysis was applied to representative models. Fig. 7 presents SHAP summary plots for XGBoost-based Dose prediction and Extra Trees–based BED prediction. Across models, AMax emerged as the most influential feature, with higher values consistently driving increased predicted outcomes. Clearance, Tmax, and AMean also contributed strongly, emphasizing the role of early uptake intensity and washout behavior. Distribution-based features such as Skewness and percentile descriptors contributed in a model-dependent manner, particularly for BED, highlighting the relevance of TAC shape. These results confirm that model predictions are driven by physiologically meaningful kinetic signatures rather than spurious correlations.

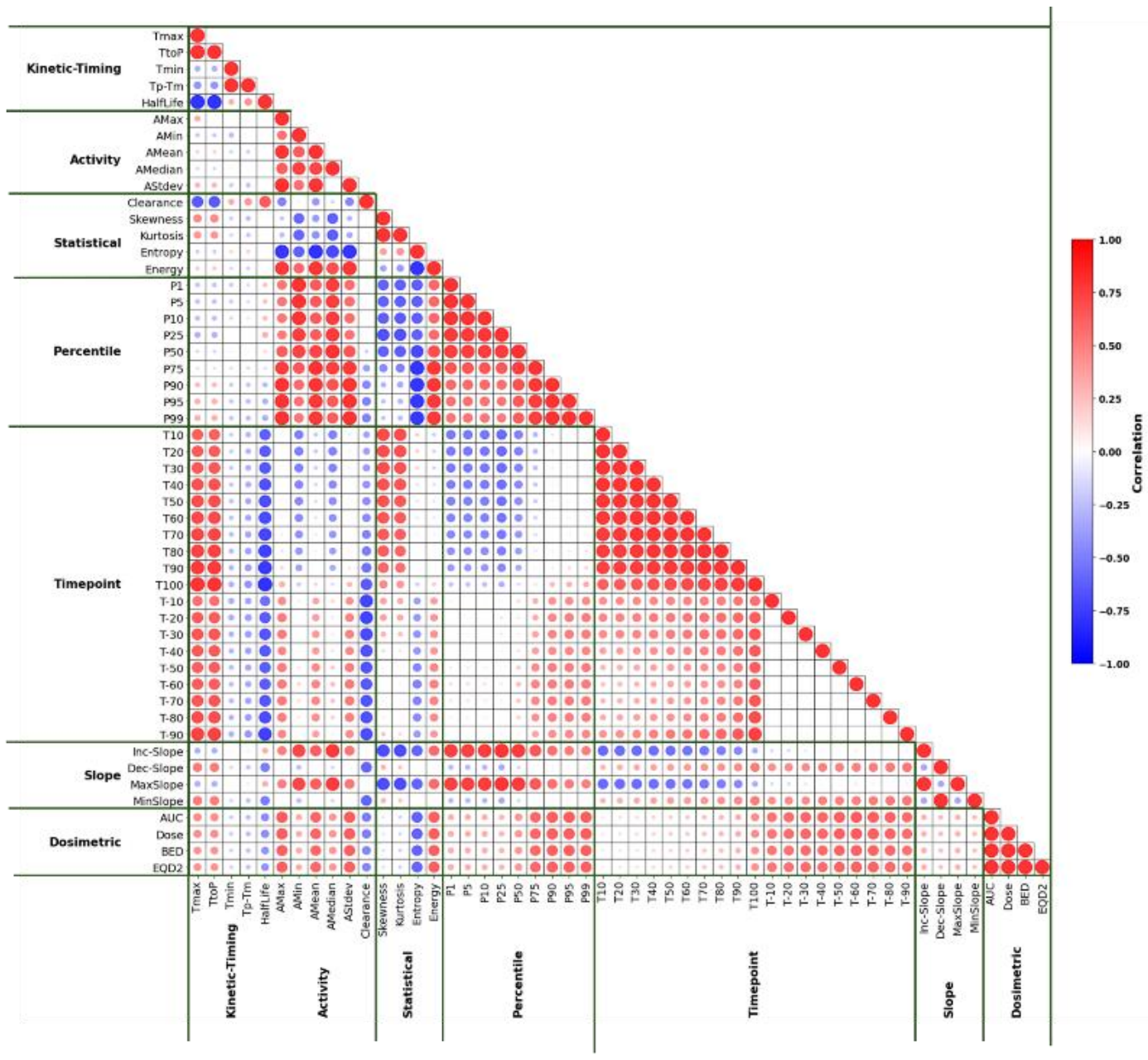

Fig. 2. Correlation map of pre-therapy dynamic PET imaging features and dosimetric outcome features (AUC, Dose, BED, and EQD2). This map illustrates the pairwise correlations between dosimetric endpoints and diverse categories of time–activity features derived from dynamic PET imaging. Features are grouped into kinetic–timing, activity, statistical descriptors, percentiles, timepoint intensities, slope-based measures, and dosimetric endpoints. Red indicates positive correlations and blue indicates negative correlations, with circle size reflecting correlation strength. The plot highlights strong interdependencies among activity-based, percentile, and dosimetric features, underscoring their shared predictive value for RPT outcome modeling

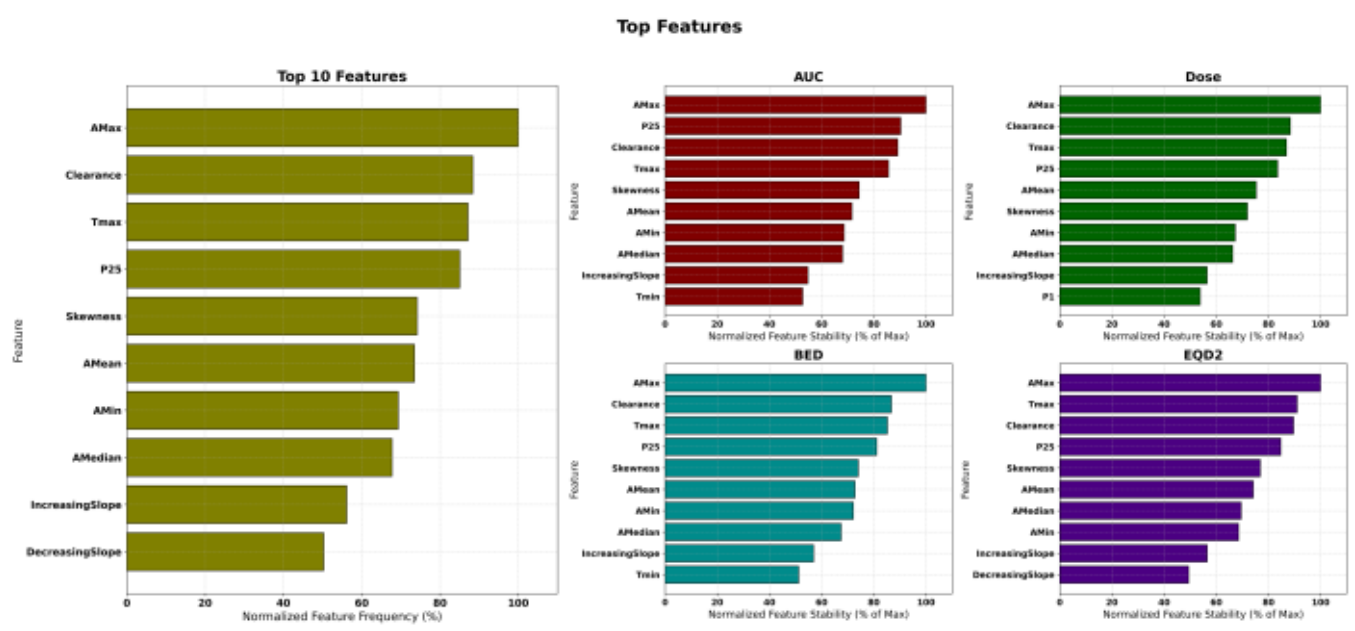

Fig. 3. Selected Top 10 Predictive Features Across All Radioisotopes, Noise Settings, and Targets. The left panel summarizes the ten most frequently selected features across all predictions, including both tumors and normal organs. These features consistently demonstrated the highest selection frequency regardless of tracer or noise setting. The right panels present the top ten predictive features ranked separately for each endpoint (AUC, Dose, BED, and EQD2). Feature stability was normalized to the maximum value within each endpoint to allow cross-comparison. Together, these plots highlight AMax, Clearance, Tmax, P25, and Skewness as highly stable predictors across multiple outcomes.

## A. *Prediction performance*

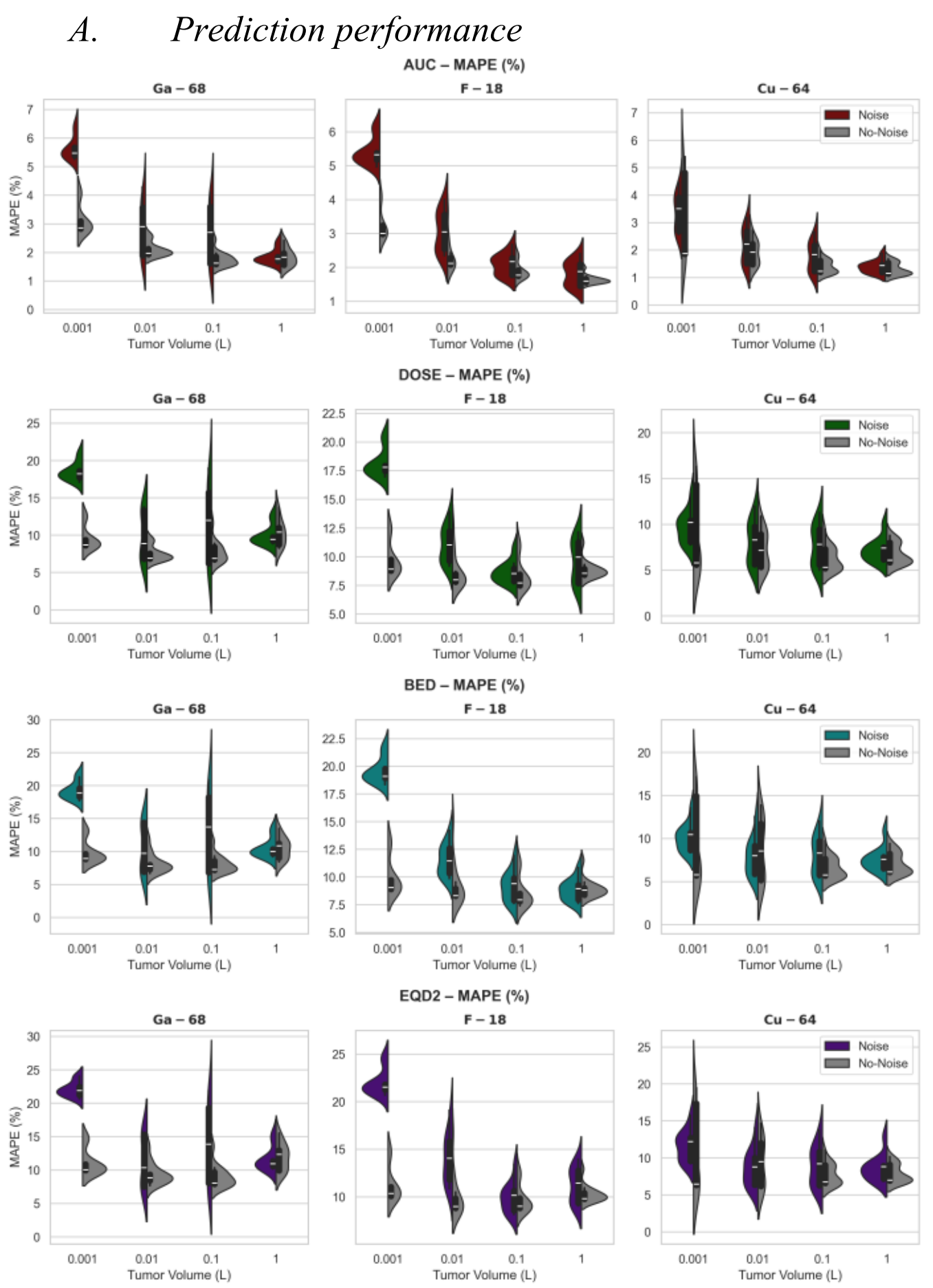

Fig. 4. Violin plots of predictive performance (MAPE%) for tumor AUC, Dose, BED, and EQD2 for $^{18}$F-DCFPYL, $^{68}$Ga-PSMA-11, and $^{64}$Cu-PSMA-617 isotopes under Noise and No-Noise conditions. $^{64}$Cu-PSMA-617 shows the most stable and lowest MAPE across volumes, while $^{18}$F-DCFPYL exhibits consistent, volume-dependent trends. $^{68}$Ga-PSMA-11 demonstrates high variability and greater noise sensitivity, especially at smaller tumor volumes.

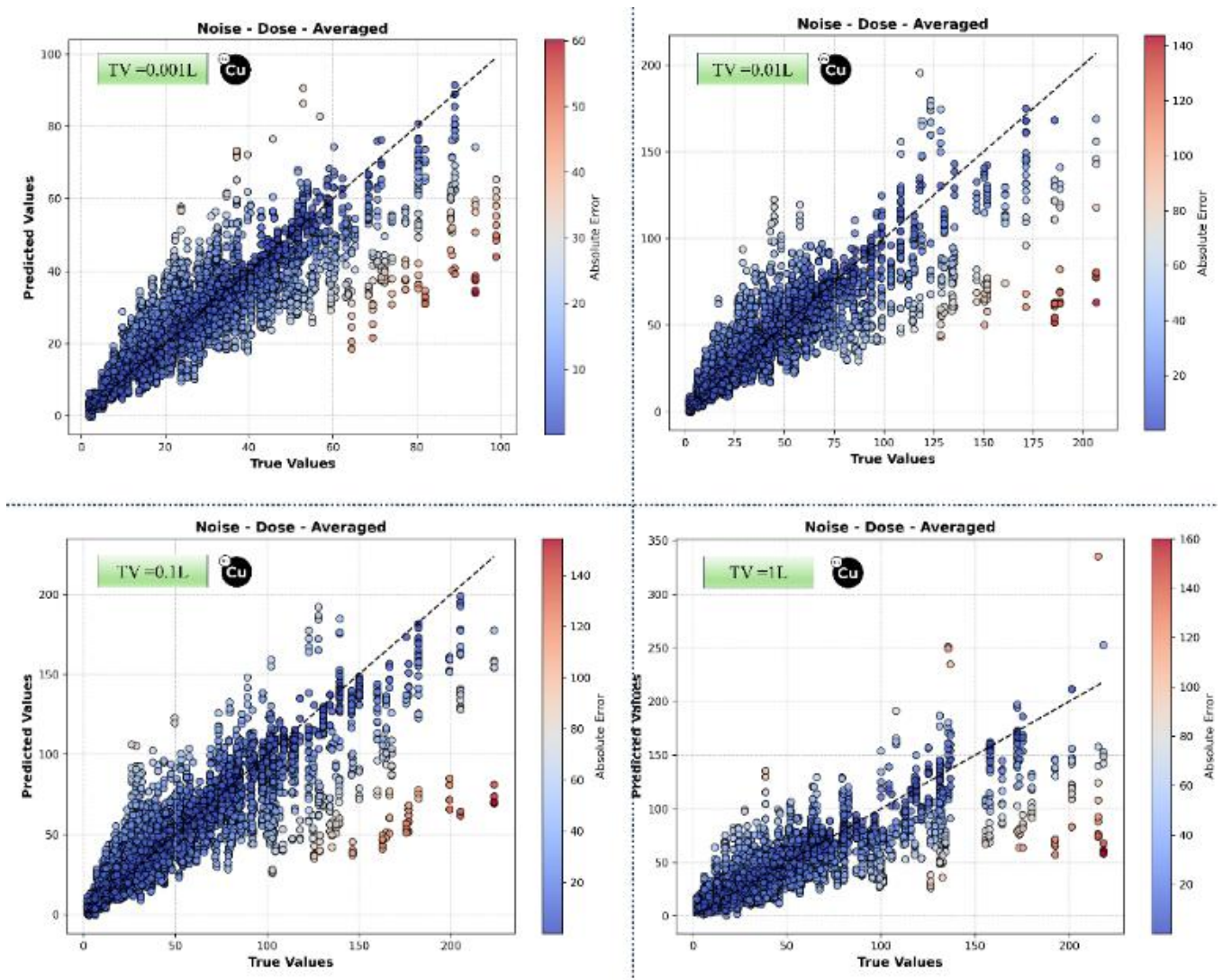

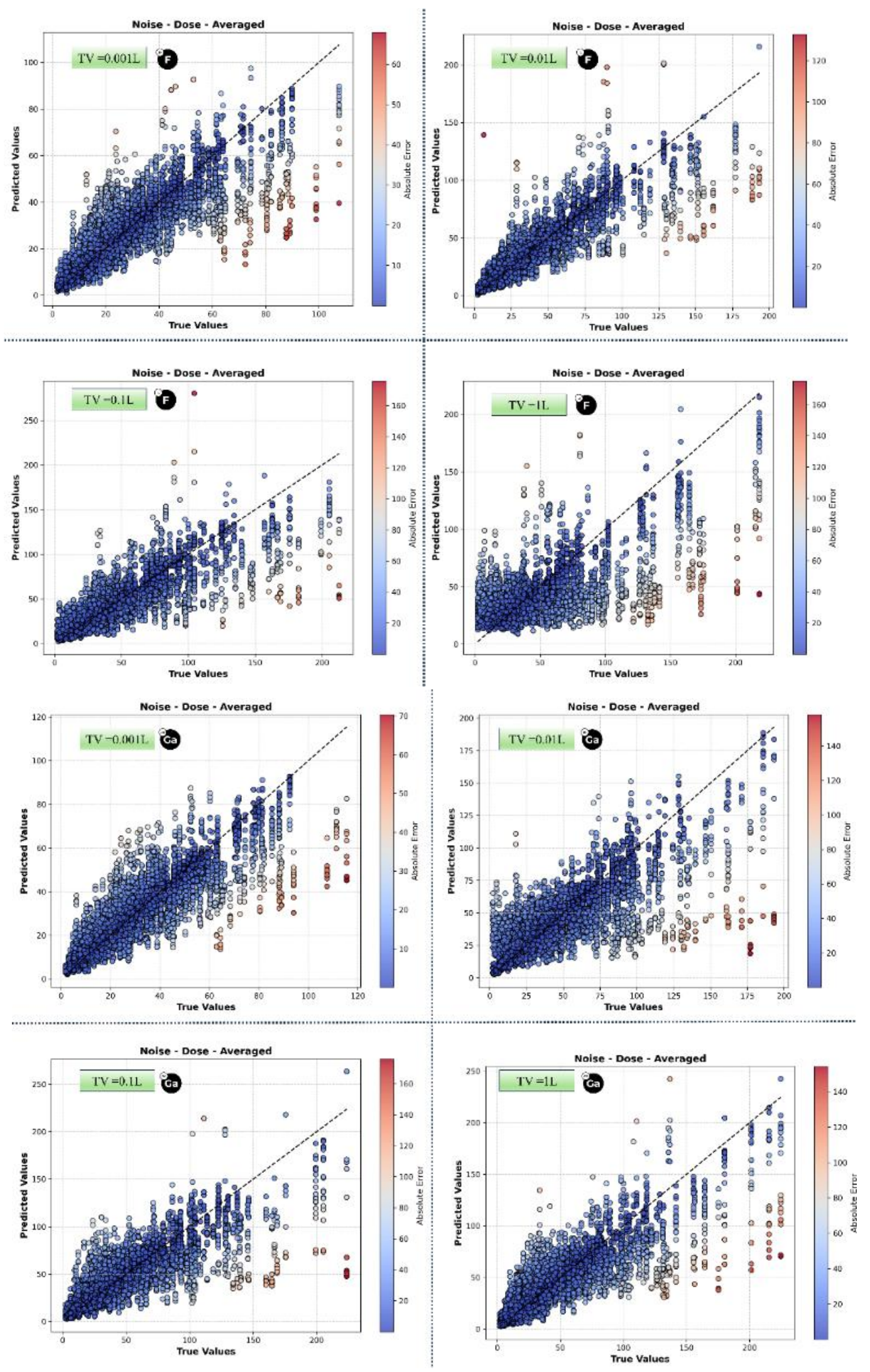

Fig. 5. Predicted vs. true absorbed dose under Noise conditions across tumor volumes (TV = 0.001–1 L) for $^{64}$Cu-PSMA-617 (top), $^{68}$Ga-PSMA-11 (middle), and $^{18}$F-DCFPyL (bottom). Points are colored by absolute error (blue = low, red = high). Accuracy improves with larger volumes, while small volumes show higher variability and error. The dashed line represents perfect agreement ($y = x$). At very small volumes, noise dominates, leading to systematic underestimation and scattered predictions, whereas for $TV \geq 0.1$ L predictions closely track the identity line, indicating more stable and clinically reliable performance

.

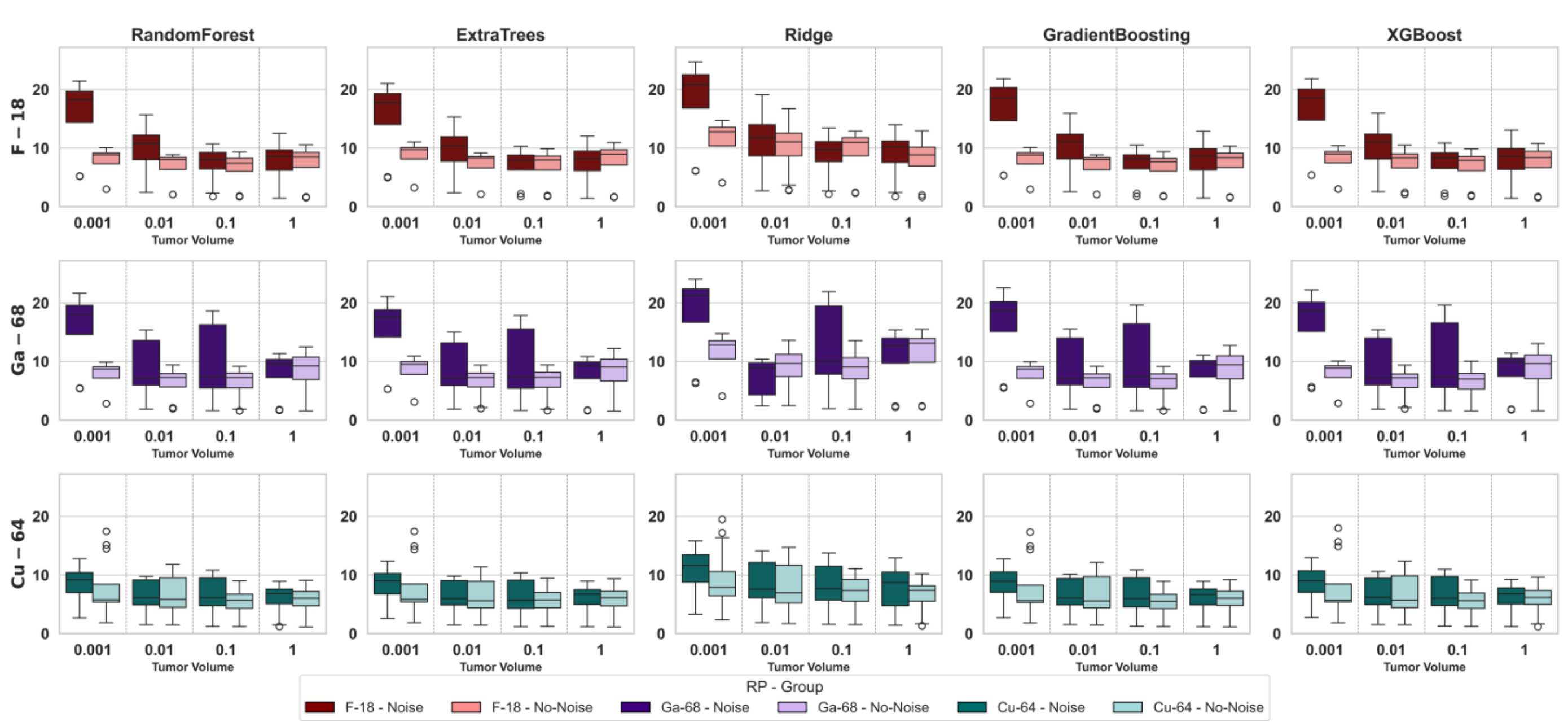

Fig. 6. Distributions of mean absolute percentage error (MAPE, %) in dose prediction across tumor volumes (0.001–1 L) for three radiopharmaceuticals ($^{18}$F-DCFPYL, $^{68}$Ga-PSMA-11, $^{64}$Cu-PSMA-617) and five machine learning models under Noise and No-Noise conditions. $^{64}$Cu-PSMA-617 consistently shows the lowest and most stable errors, particularly with Ridge and XGBoost. In contrast,$^{68}$Ga-PSMA-11 demonstrates the greatest sensitivity to noise, with error spread decreasing at larger volumes. $^{18}$F-DCFPYL models perform intermediately, with gradual error reduction as volume increases. Overall, predictive accuracy improves with tumor volume, highlighting reduced variability and stronger robustness in larger targets.

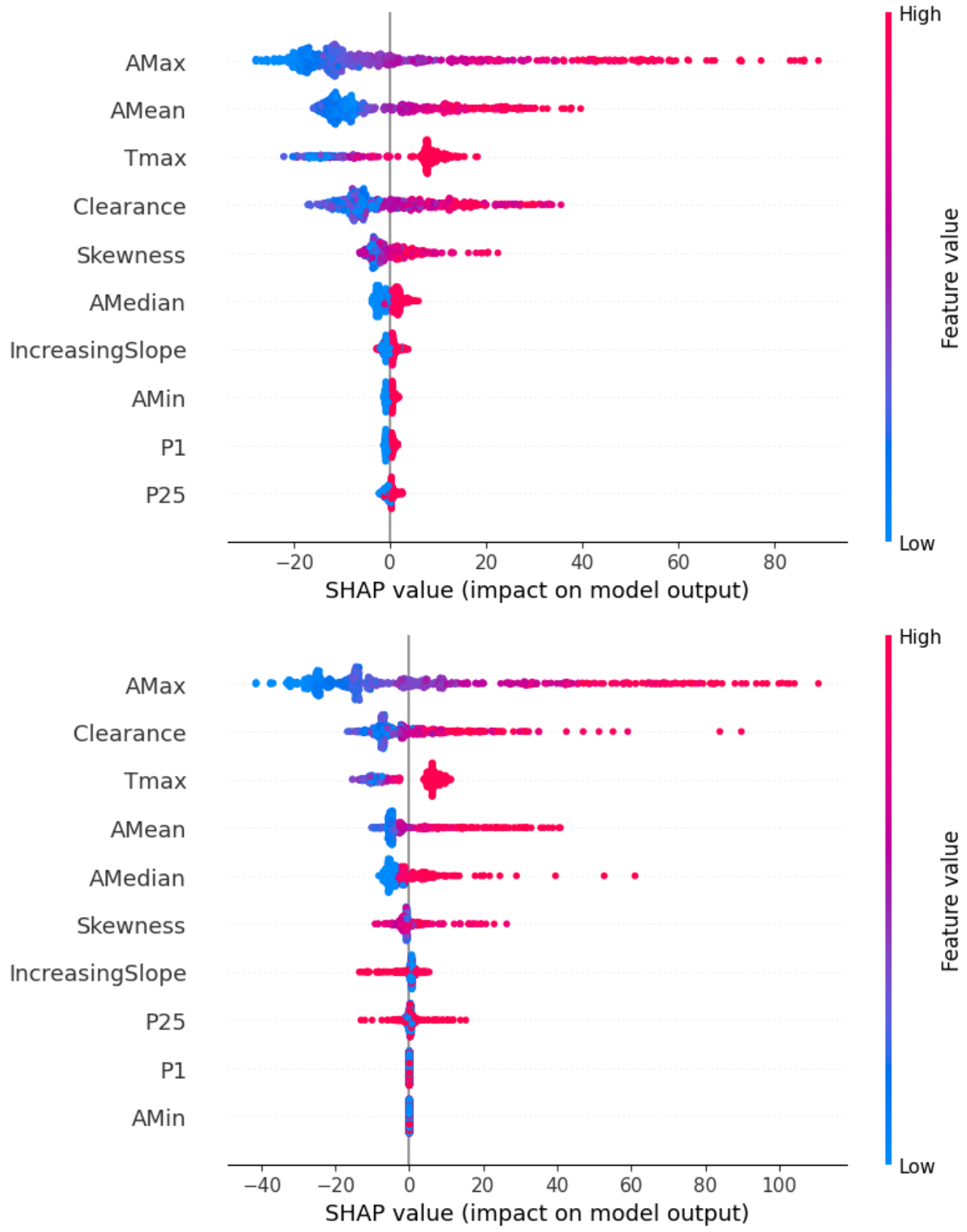

Fig. 7. SHAP analysis for interpretability of predictive machine learning models. Shown are SHAP summary plots illustrating the contribution of the top predictive features to model output. The upper panel corresponds to XGBoost predictions for Dose, while the lower panel corresponds to ExtraTrees predictions for BED at a tumor volume of 1 L. Each dot represents a single sample, with color indicating the magnitude of the feature value (blue = low, red = high) and the position on the x-axis indicating the SHAP value (impact on prediction). Features such as AMax, AMean, Tmax, and Clearance emerged as the most influential predictors, with consistent positive SHAP values at higher feature levels, highlighting their critical role in dosimetry outcome prediction.

## IV. DISCUSSION

This study demonstrates that predictive dosimetry derived from pre-therapeutic PET imaging is feasible and robust within a virtual theranostic trial framework, supporting its potential role in personalized radiopharmaceutical therapy (RPT). By leveraging dynamic PET time–activity curves (TACs) and TAC-derived kinetic features, the proposed framework captures inter-patient variability in radiopharmaceutical uptake, clearance, and organ-specific biodistribution. These imaging-derived descriptors enabled accurate prediction of both physical (AUC, absorbed dose) and biologically weighted (BED, EQD2) endpoints across tumor volumes and normal organs. This capability directly addresses a major limitation of current clinical practice, where fixed or weight-based administered activities fail to account for substantial heterogeneity in tumor burden, PSMA expression, and normal-organ uptake, potentially resulting in clinically meaningful under- or overtreatment [43–46].

Across all evaluated conditions, $^{64}$Cu-PSMA-617 PET imaging consistently yielded the most accurate and noise-robust predictions. As shown in Table III and corresponding normal-organ results, $^{64}$Cu-based imaging produced lower MAPE values and narrower error distributions for absorbed dose, BED, and EQD2 compared with $^{18}$F-DCFPyL and $^{68}$Ga-PSMA-11, particularly under noisy conditions and for small tumor volumes. This advantage is primarily attributable to the longer physical half-life of $^{64}$Cu (~12.7 h), which enables extended imaging windows and smoother TACs, thereby stabilizing kinetic feature extraction. In contrast, the short half-life and rapid kinetics of $^{68}$Ga (~68 min) amplify sensitivity to temporal sampling limitations and noise, resulting in broader error distributions and reduced discriminative power, particularly for biologically weighted endpoints. $^{18}$F-DCFPyL exhibited intermediate behavior, with predictable volume-dependent improvements but greater noise sensitivity at small volumes. Although these tracer-specific differences were identified under controlled simulation conditions rather than clinical acquisitions, they highlight an important practical implication: tracers with longer half-lives and smoother kinetics may be intrinsically more resilient to acquisition noise and sparse sampling in predictive dosimetry pipelines.

Tumor volume emerged as a dominant determinant of prediction reliability. Across all tracers, endpoints, and models, prediction errors were highest for small lesions (≤0.01 L) and decreased systematically with increasing tumor volume, as reflected by reduced MAPE and tighter distributions (Figs. 4–5). This trend is consistent with fundamental imaging limitations, including reduced count statistics, partial-volume–like effects, and increased relative noise in small lesions. Clinically, these findings suggest that predictive dosimetry for very small tumor deposits should be interpreted with caution and may benefit from uncertainty-aware modeling or optimized acquisition protocols.

Although tumors are smaller than most normal organs, their predictive performance was consistently superior across all endpoints, which can be explained by both PBPK-driven kinetics and dosimetric physics. In the PBPK model, tumors are characterized by higher PSMA receptor density, stronger ligand internalization, and more focal activity retention, resulting in TACs with higher contrast and more informative kinetic features. From a dosimetric standpoint, the smaller tumor mass leads to larger S-values, such that a given time-integrated activity translates into a higher absorbed dose per unit activity compared with normal organs. This amplification strengthens the coupling between PET-derived kinetics and dose-related targets, thereby improving learnability and reducing relative prediction error. In contrast, normal organs exhibit larger volumes, lower receptor specificity, and greater physiologic dispersion, which dilute activity–dose coupling and increase variability, ultimately degrading predictive accuracy despite their larger size

Feature-level analysis further supports the physiological plausibility of the predictive framework. Across organs and endpoints, features describing uptake magnitude and kinetic behavior, particularly AMax, Clearance, Tmax, and early distribution metrics were consistently prioritized during model training. These findings align with established relationships between PET-derived uptake metrics, time-integrated activity, absorbed dose, and biological effect in radionuclide therapy [47]. Feature selection was performed independently for each organ and endpoint, with highly correlated variables removed (Pearson $r > 0.90$), ensuring model stability and minimizing collinearity. The convergence of selected features across folds and models indicates robust and reproducible feature importance rankings rather than overfitting to simulation artifacts.

Model performance was strongly influenced by algorithmic robustness under noisy conditions. Regularized approaches such as Ridge regression and gradient-boosted ensembles (XGBoost) consistently demonstrated lower error variability and greater noise resilience than less constrained ensemble methods, particularly for small tumors and BED/EQD2 prediction. These results emphasize that physiologically meaningful feature engineering alone is insufficient; noise-tolerant learning strategies are essential for reliable deployment of AI-driven predictive dosimetry in clinically realistic settings.

The use of PBPK modeling to generate paired PET–RPT datasets constitutes a major strength of this study. The PBPK framework enabled systematic exploration of radiopharmaceutical kinetics, tumor volumes, organ-specific uptake, and noise effects within a controlled environment, facilitating large-scale training and benchmarking of predictive models that would be impractical using clinical data alone. This approach also provides a natural foundation for TDT development, wherein patient-specific physiological parameters can be adjusted to generate individualized predictions and virtual treatment scenarios [15,31,48].

Several limitations should be acknowledged. First, the simulated PET data included extended imaging time points that exceed routine clinical protocols and may overestimate the precision achievable in practice. Second, due to limited availability of comprehensive agent-specific *in vivo* data, kinetic parameters were harmonized across tracers, potentially underrepresenting true inter-tracer variability. Third, while absorbed dose was directly derived from the PBPK model, BED and EQD2 depend on radiobiological parameters adopted from the literature and applied in a population-averaged manner, introducing uncertainty that propagates into biological dose estimates. Finally, although the framework was rigorously evaluated in a large synthetic cohort, external validation using clinical post-therapy dosimetry data remains necessary before clinical translation.

TABLE III. PREDICTIVE PERFORMANCE (MAPE%) FOR NORMAL ORGANS OVER DIFFERENT TARGETS AND RADIOISOTOPES AND OVER DIFFERENT TUMOR VOLUMES (0.001, 0.01, 0.1, 1 L)

| | Organ | RP | 0.001 | | 0.01 | | 0.1 | | 1 | |
|---|---|---|---|---|---|---|---|---|---|---|
| | | | Noise | No Noise | Noise | No Noise | Noise | No Noise | Noise | No Noise |
| BED | Kidney | $^{18}$F-DCFPYL | 15.4 ± 4.3 | 5.2 ± 8.0 | 15.6 ± 4.4 | 6.1 ± 7.8 | 16.6 ± 4.3 | 10.0 ± 6.1 | 18.4 ± 3.7 | 11.7 ± 4.9 |
| | | $^{68}$Ga-PSMA-11 | 17.5 ± 2.7 | 5.2 ± 7.8 | 17.4 ± 2.3 | 5.9 ± 6.9 | 17.1 ± 1.9 | 10.2 ± 5.5 | 19.3 ± 2.2 | 13.6 ± 5.5 |
| | | $^{64}$Cu-PSMA-617 | 9.2 ± 6.7 | 4.0 ± 6.4 | 7.9 ± 7.4 | 7.0 ± 7.2 | 12.0 ± 8.0 | 4.9 ± 5.7 | 16.2 ± 8.7 | 7.2 ± 5.5 |
| | Salivary gland | $^{18}$F-DCFPYL | 23.3 ± 2.3 | 12.3 ± 10.5 | 27.4 ± 6.8 | 12.2 ± 9.5 | 27.9 ± 7.1 | 14.6 ± 6.6 | 25.6 ± 2.9 | 16.2 ± 4.5 |
| | | $^{68}$Ga-PSMA-11 | 28.3 ± 4.9 | 13.1 ± 11.3 | 23.1 ± 1.5 | 13.4 ± 10.4 | 21.9 ± 1.4 | 16.5 ± 6.4 | 25.4 ± 1.9 | 18.3 ± 5.5 |
| | | $^{64}$Cu-PSMA-617 | 20.1 ± 5.5 | 9.6 ± 9.2 | 22.2 ± 7.1 | 10.3 ± 8.3 | 15.6 ± 1.6 | 10.6 ± 8.5 | 20.0 ± 2.4 | 8.7 ± 5.9 |
| | Liver | $^{18}$F-DCFPYL | 42.2 ± 19.0 | 23.2 ± 17.9 | 37.3 ± 16.5 | 26.1 ± 13.5 | 35.5 ± 14.5 | 32.3 ± 6.8 | 23.9 ± 2.9 | 21.3 ± 5.5 |
| | | $^{68}$Ga-PSMA-11 | 42.3 ± 11.7 | 15.0 ± 17.1 | 42.5 ± 14.0 | 18.4 ± 16.4 | 39.8 ± 11.2 | 28.8 ± 15.7 | 30.6 ± 3.8 | 23.0 ± 7.4 |
| | | $^{64}$Cu-PSMA-617 | 17.9 ± 4.0 | 7.5 ± 7.6 | 17.5 ± 3.9 | 10.9 ± 6.6 | 17.8 ± 4.3 | 11.1 ± 6.4 | 26.4 ± 6.8 | 19.0 ± 5.9 |
| | Spleen | $^{18}$F-DCFPYL | 31.1 ± 4.2 | 9.0 ± 10.2 | 31.8 ± 4.8 | 10.4 ± 9.8 | 30.6 ± 2.3 | 13.6 ± 6.3 | 34.2 ± 4.8 | 19.6 ± 4.3 |
| | | $^{68}$Ga-PSMA-11 | 44.9 ± 5.5 | 16.0 ± 22.7 | 43.0 ± 4.6 | 11.2 ± 12.6 | 40.2 ± 4.1 | 18.8 ± 6.1 | 22.6 ± 12.6 | 11.1 ± 9.5 |
| | | $^{64}$Cu-PSMA-617 | 34.0 ± 5.5 | 11.7 ± 11.0 | 32.5 ± 4.1 | 13.0 ± 10.5 | 32.0 ± 5.6 | 15.6 ± 8.8 | 37.3 ± 6.0 | 18.8 ± 6.3 |
| | Red marrow | $^{18}$F-DCFPYL | 18.7 ± 6.7 | 4.0 ± 5.5 | 16.5 ± 5.2 | 4.7 ± 5.8 | 21.6 ± 9.9 | 6.6 ± 7.1 | 14.4 ± 3.5 | 7.4 ± 2.9 |
| | | $^{68}$Ga-PSMA-11 | 29.4 ± 3.4 | 7.4 ± 13.0 | 29.7 ± 4.1 | 8.3 ± 12.5 | 27.0 ± 5.9 | 16.4 ± 14.6 | 16.1 ± 2.6 | 12.8 ± 7.4 |

| | | $^{64}$Cu-PSMA-617 | 11.6 ± 5.0 | 3.0 ± 4.0 | 12.3 ± 5.8 | 8.5 ± 7.0 | 11.2 ± 4.2 | 4.2 ± 4.5 | 14.0 ± 3.9 | 10.0 ± 7.1 |
|---|---|---|---|---|---|---|---|---|---|---|

| EQD2 | **Organ** | **RP** | **0.001** | | **0.01** | | **0.1** | | **1** | |
|---|---|---|---|---|---|---|---|---|---|---|
| | | | **Noise** | **No Noise** | **Noise** | **No Noise** | **Noise** | **No Noise** | **Noise** | **No Noise** |
| | **Kidney** | **$^{18}$F-DCFPYL** | 19.5 ± 6.8 | 6.2 ± 9.5 | 19.4 ± 6.7 | 7.2 ± 9.2 | 19.2 ± 4.8 | 11.1 ± 7.1 | 21.4 ± 4.7 | 14.4 ± 5.7 |
| | | **$^{68}$Ga-PSMA-11** | 21.0 ± 4.5 | 6.0 ± 8.9 | 20.7 ± 3.1 | 6.6 ± 7.5 | 20.1 ± 3.3 | 11.7 ± 6.3 | 21.9 ± 3.6 | 16.4 ± 5.9 |
| | | **$^{64}$Cu-PSMA-617** | 7.8 ± 6.1 | 4.8 ± 7.6 | 13.2 ± 10.5 | 8.5 ± 7.0 | 13.8 ± 10.1 | 6.1 ± 7.1 | 20.4 ± 12.3 | 8.3 ± 6.7 |
| | **Salivary gland** | **$^{18}$F-DCFPYL** | 23.9 ± 1.8 | 13.7 ± 11.8 | 30.1 ± 8.0 | 13.4 ± 10.4 | 32.2 ± 8.8 | 15.9 ± 7.5 | 27.7 ± 3.6 | 17.5 ± 5.3 |
| | | **$^{68}$Ga-PSMA-11** | 29.3 ± 4.2 | 14.2 ± 12.5 | 25.0 ± 1.3 | 14.8 ± 11.6 | 23.2 ± 2.0 | 18.1 ± 7.0 | 27.4 ± 1.8 | 20.3 ± 5.7 |
| | | **$^{64}$Cu-PSMA-617** | 23.4 ± 7.3 | 10.3 ± 9.7 | 23.9 ± 7.6 | 13.0 ± 8.0 | 22.4 ± 6.5 | 11.5 ± 9.1 | 24.4 ± 5.1 | 9.9 ± 6.9 |
| **EQD2** | **Liver** | **$^{18}$F-DCFPYL** | 53.4 ± 24.7 | 30.4 ± 23.2 | 50.0 ± 24.2 | 34.3 ± 19.6 | 44.1 ± 18.4 | 40.4 ± 9.5 | 29.0 ± 4.2 | 27.4 ± 6.7 |
| | | **$^{68}$Ga-PSMA-11** | 53.6 ± 18.3 | 16.7 ± 23.5 | 51.6 ± 15.4 | 20.8 ± 21.8 | 49.7 ± 14.3 | 34.5 ± 17.6 | 36.9 ± 5.1 | 27.1 ± 8.4 |
| | | **$^{64}$Cu-PSMA-617** | 22.1 ± 6.3 | 9.8 ± 11.7 | 23.6 ± 6.1 | 14.3 ± 8.7 | 24.0 ± 6.7 | 14.8 ± 9.2 | 32.0 ± 8.9 | 22.9 ± 8.4 |
| | **Spleen** | **$^{18}$F-DCFPYL** | 35.1 ± 5.0 | 10.5 ± 11.4 | 33.2 ± 4.7 | 11.0 ± 10.0 | 33.0 ± 4.2 | 16.7 ± 8.0 | 35.8 ± 5.5 | 21.8 ± 7.6 |
| | | **$^{68}$Ga-PSMA-11** | 46.0 ± 5.9 | 16.3 ± 23.1 | 48.2 ± 5.8 | 11.0 ± 12.4 | 41.1 ± 4.6 | 19.2 ± 6.4 | 24.4 ± 13.8 | 12.5 ± 9.4 |
| | | **$^{64}$Cu-PSMA-617** | 35.1 ± 5.7 | 13.5 ± 12.6 | 36.9 ± 7.4 | 25.0 ± 8.0 | 33.8 ± 6.4 | 17.4 ± 10.0 | 42.6 ± 7.4 | 20.1 ± 7.4 |
| | **Red marrow** | **$^{18}$F-DCFPYL** | 19.5 ± 6.8 | 5.4 ± 6.2 | 19.5 ± 6.8 | 5.8 ± 6.4 | 21.3 ± 9.3 | 7.4 ± 7.4 | 14.7 ± 3.5 | 8.5 ± 3.5 |
| | | **$^{68}$Ga-PSMA-11** | 32.2 ± 4.3 | 7.6 ± 13.4 | 30.7 ± 3.8 | 8.3 ± 12.5 | 27.4 ± 4.8 | 17.5 ± 15.6 | 17.4 ± 2.8 | 13.0 ± 7.2 |
| | | **$^{64}$Cu-PSMA-617** | 11.2 ± 4.9 | 3.4 ± 4.0 | 12.6 ± 5.8 | 12.0 ± 7.0 | 12.7 ± 5.1 | 14.0 ± 13.8 | 15.6 ± 4.7 | 10.7 ± 7.3 |

| | **Organ** | **RP** | **0.001** | | **0.01** | | **0.1** | | **1** | |
|---|---|---|---|---|---|---|---|---|---|---|
| | | | **Noise** | **No Noise** | **Noise** | **No Noise** | **Noise** | **No Noise** | **Noise** | **No Noise** |
| | **Kidney** | **$^{18}$F-DCFPYL** | 1.9 ± 0.5 | 0.70 ± 1.0 | 1.9 ± 0.40 | 0.7 ± 0.9 | 2.0 ± 0.4 | 1.2 ± 0.7 | 2.1 ± 0.4 | 1.6 ± 0.7 |
| | | **$^{68}$Ga-PSMA-11** | 2.1 ± 0.3 | 0.7 ± 1.0 | 2.2 ± 0.2 | 0.8 ± 0.9 | 2.2 ± 0.2 | 1.3 ± 0.7 | 2.2 ± 0.2 | 1.7 ± 0.7 |
| | | **$^{64}$Cu-PSMA-617** | 1.3 ± 0.8 | 0.5 ± 0.7 | 1.2 ± 0.8 | 0.8 ± 0.7 | 1.3 ± 0.7 | 0.6 ± 0.6 | 1.8 ± 0.8 | 0.8 ± 0.6 |
| | **Salivary gland** | **$^{18}$F-DCFPYL** | 3.3 ± 0.2 | 1.7 ± 1.5 | 3.7 ± 0.7 | 1.8 ± 1.4 | 3.6 ± 0.4 | 2.1 ± 0.9 | 3.4 ± 0.5 | 2.4 ± 0.6 |
| | | **$^{68}$Ga-PSMA-11** | 3.6 ± 0.4 | 1.9 ± 1.6 | 3.3 ± 0.2 | 2.0 ± 1.5 | 3.1 ± 0.3 | 2.5 ± 0.9 | 3.3 ± 0.2 | 2.7 ± 0.8 |
| | | **$^{64}$Cu-PSMA-617** | 2.9 ± 0.7 | 1.4 ± 1.3 | 3.0 ± 0.8 | 2.0 ± 1.1 | 3.1 ± 0.9 | 1.6 ± 1.2 | 2.8 ± 0.3 | 1.1 ± 0.7 |
| **AUC** | **Liver** | **$^{18}$F-DCFPYL** | 2.4 ± 0.9 | 1.5 ± 1.1 | 2.4 ± 0.8 | 1.7 ± 0.8 | 2.2 ± 0.7 | 2.1 ± 0.4 | 1.7 ± 0.2 | 1.4 ± 0.3 |
| | | **$^{68}$Ga-PSMA-11** | 2.6 ± 0.6 | 1.1 ± 1.1 | 2.4 ± 0.6 | 1.0 ± 0.9 | 2.5 ± 0.6 | 1.7 ± 0.7 | 2.0 ± 0.2 | 1.5 ± 0.5 |
| | | **$^{64}$Cu-PSMA-617** | 1.2 ± 0.3 | 0.5 ± 0.6 | 1.2 ± 0.2 | 0.8 ± 0.4 | 1.3 ± 0.3 | 0.8 ± 0.5 | 1.7 ± 0.4 | 1.3 ± 0.4 |
| | **Spleen** | **$^{18}$F-DCFPYL** | 2.6 ± 0.2 | 0.8 ± 0.9 | 2.8 ± 0.2 | 0.9 ± 0.9 | 2.7 ± 0.2 | 1.2 ± 0.7 | 2.9 ± 0.2 | 1.9 ± 0.5 |
| | | **$^{68}$Ga-PSMA-11** | 3.4 ± 0.3 | 2.2 ± 2.1 | 3.6 ± 0.3 | 1.1 ± 1.1 | 3.3 ± 0.2 | 1.8 ± 0.7 | 1.9 ± 1.0 | 0.9 ± 0.7 |
| | | **$^{64}$Cu-PSMA-617** | 3.0 ± 0.4 | 0.9 ± 0.8 | 3.1 ± 0.4 | 1.3 ± 1.0 | 3.1 ± 0.5 | 1.3 ± 0.7 | 3.1 ± 0.3 | 1.7 ± 0.5 |
| | **Red marrow** | **$^{18}$F-DCFPYL** | 1.4 ± 0.5 | 0.3 ± 0.4 | 1.6 ± 0.6 | 0.3 ± 0.4 | 1.5 ± 0.6 | 0.5 ± 0.5 | 1.3 ± 0.4 | 0.6 ± 0.2 |
| | | **$^{68}$Ga-PSMA-11** | 2.4 ± 0.2 | 0.6 ± 1.0 | 2.3 ± 0.3 | 0.6 ± 0.9 | 2.1 ± 0.4 | 1.3 ± 1.1 | 1.4 ± 0.2 | 1.0 ± 0.5 |
| | | **$^{64}$Cu-PSMA-617** | 0.8 ± 0.4 | 1.1 ± 1.3 | 0.9 ± 0.4 | 1.0 ± 0.9 | 0.8 ± 0.3 | 1.0 ± 1.0 | 1.1 ± 0.3 | 0.8 ± 0.6 |
| **Dose** | **Organ** | **RP** | **0.001** | | **0.01** | | **0.1** | | **1** | |
| | | | **Noise** | **No Noise** | **Noise** | **No Noise** | **Noise** | **No Noise** | **Noise** | **No Noise** |
| | **Kidney** | **$^{18}$F-DCFPYL** | 16.0 ± 5.1 | 5.3 ± 8.0 | 14.7 ± 4.2 | 5.8 ± 7.3 | 16.3 ± 3.8 | 9.7 ± 5.9 | 19.9 ± 5.4 | 13.2 ± 5.9 |
| | | **$^{68}$Ga-PSMA-11** | 16.8 ± 1.9 | 5.0 ± 7.5 | 16.8 ± 1.9 | 5.7 ± 6.6 | 16.9 ± 1.6 | 9.8 ± 5.4 | 18.2 ± 2.2 | 13.9 ± 4.9 |
| | | **$^{64}$Cu-PSMA-617** | 6.9 ± 4.3 | 4.0 ± 6.3 | 9.8 ± 6.8 | 4.3 ± 6.3 | 11.3 ± 7.6 | 4.9 ± 5.8 | 16.8 ± 9.6 | 6.9 ± 5.4 |
| | **Salivary gland** | **$^{18}$F-DCFPYL** | 21.5 ± 1.7 | 12.3 ± 10.7 | 26.5 ± 6.4 | 12.2 ± 9.5 | 27.2 ± 7.0 | 14.3 ± 6.3 | 27.5 ± 5.7 | 16.5 ± 4.2 |
| | | **$^{68}$Ga-PSMA-11** | 24.8 ± 2.7 | 12.9 ± 11.4 | 22.2 ± 1.4 | 13.5 ± 10.7 | 20.8 ± 1.7 | 16.6 ± 6.5 | 24.7 ± 1.7 | 18.9 ± 5.6 |
| | | **$^{64}$Cu-PSMA-617** | 19.6 ± 5.0 | 9.5 ± 9.1 | 22.4 ± 7.1 | 10.3 ± 8.5 | 15.8 ± 1.8 | 10.2 ± 8.2 | 20.3 ± 3.0 | 8.8 ± 6.3 |
| | **Liver** | **$^{18}$F-DCFPYL** | 30.6 ± 12.9 | 17.9 ± 13.6 | 29.2 ± 11.9 | 20.7 ± 10.8 | 28.0 ± 11.5 | 25.1 ± 5.0 | 19.3 ± 2.0 | 16.1 ± 3.2 |
| | | **$^{68}$Ga-PSMA-11** | 31.6 ± 9.0 | 15.7 ± 14.6 | 31.3 ± 8.8 | 15.6 ± 13.1 | 30.6 ± 7.7 | 20.5 ± 9.7 | 24.8 ± 3.0 | 18.5 ± 5.9 |
| | | **$^{64}$Cu-PSMA-617** | 13.8 ± 3.0 | 4.7 ± 4.6 | 14.1 ± 2.8 | 8.1 ± 4.9 | 14.4 ± 2.6 | 8.6 ± 5.1 | 20.8 ± 5.1 | 15.2 ± 4.7 |
| | **Spleen** | **$^{18}$F-DCFPYL** | 27.7 ± 2.5 | 8.0 ± 8.7 | 27.4 ± 2.4 | 8.9 ± 7.9 | 28.6 ± 2.5 | 12.5 ± 5.3 | 30.6 ± 3.7 | 18.8 ± 5.1 |
| | | **$^{68}$Ga-PSMA-11** | 37.2 ± 3.8 | 13.9 ± 19.5 | 39.4 ± 4.0 | 9.8 ± 10.2 | 36.8 ± 4.0 | 16.1 ± 5.7 | 21.0 ± 11.3 | 10.4 ± 8.5 |
| | | **$^{64}$Cu-PSMA-617** | 28.9 ± 3.3 | 10.0 ± 9.4 | 29.0 ± 2.4 | 11.7 ± 8.5 | 29.7 ± 4.3 | 14.2 ± 7.0 | 33.5 ± 5.9 | 17.7 ± 6.0 |
| | **Red marrow** | **$^{18}$F-DCFPYL** | 18.4 ± 7.1 | 4.3 ± 5.7 | 14.5 ± 4.7 | 4.2 ± 5.1 | 19.2 ± 7.4 | 6.2 ± 6.5 | 13.7 ± 2.9 | 7.5 ± 2.7 |
| | | **$^{68}$Ga-PSMA-11** | 28.9 ± 3.3 | 7.0 ± 12.3 | 28.6 ± 3.6 | 8.0 ± 12.2 | 25.4 ± 4.3 | 15.3 ± 13.6 | 16.2 ± 2.2 | 11.3 ± 5.8 |
| | | **$^{64}$Cu-PSMA-617** | 10.2 ± 4.6 | 2.9 ± 3.7 | 11.4 ± 5.3 | 3.8 ± 4.0 | 10.4 ± 4.4 | 12.1 ± 12.1 | 13.1 ± 3.6 | 10.4 ± 7.6 |

## V. CONCLUSION

This study demonstrates the feasibility and clinical potential of integrating physiologically based pharmacokinetic (PBPK) modeling with machine learning for predictive dosimetry in radiopharmaceutical therapy. By leveraging dynamic PET–derived time-activity features and theranostic RPT data, we developed a machine-learning framework capable of predicting both physical (AUC, Dose) and biologically weighted (BED, EQD2) dosimetric endpoints across tumor volumes and normal organs. Among the evaluated PET tracers, $^{64}$Cu-based imaging exhibited the most robust and noise-tolerant performance in this controlled simulation setting, likely reflecting its favorable physical half-life and kinetic characteristics, whereas $^{68}$Ga showed greater variability and $^{18}$F demonstrated volume-dependent behavior. These tracer-specific trends, observed under idealized and standardized conditions, highlight the potential influence of tracer selection on predictive dosimetry performance under practical acquisition constraints, while underscoring the need for future validation using real-world clinical data. The identification of physiologically meaningful kinetic features, combined with noise-aware machine-learning strategies, proved essential for reducing prediction variability, particularly in small-volume tumors. Overall, this simulation-based framework enables large-scale, controlled evaluation of predictive dosimetry models and provides a strong methodological foundation for the development of digital twins and data-driven, patient-specific dosimetry workflows, supporting the continued advancement of personalized theranostic strategies.

## DATA AVAILABILITY

The datasets generated and analyzed in the study are available from the corresponding author upon reasonable request.

## CODE AVAILABILITY

The code developed for data simulation, feature extraction, and model training is available on GitHub at: https://github.com/habdollahin/PBPK_TDT

## ACKNOWLEDGEMENTS

We acknowledge funding support from the Canadian Institutes of Health Research (CIHR) Project Grants PJT-180251 and PJT-197861. The authors would like to express their sincere gratitude to Pedro Esquinas, Maziar Sabouri, and Omid Gharibi for their valuable insights and constructive discussions.